\documentclass[journal]{IEEEtran}
\usepackage{times}
\usepackage{epsfig}
\usepackage[utf8]{inputenc}
\usepackage{graphicx}
\usepackage{amsmath}
\usepackage{amssymb}
\usepackage{array}
\usepackage[justification=centering]{caption}

\begin{document}

\title{SSS-PRNU: Privacy-Preserving PRNU Based Camera Attribution using Shamir Secret Sharing}



\author{
    \IEEEauthorblockN{Riyanka Jena\IEEEauthorrefmark{1}, Priyanka Singh\IEEEauthorrefmark{1}, Manoranjan Mohanty\IEEEauthorrefmark{2}}
    \IEEEauthorblockA{\\\IEEEauthorrefmark{1}
   Dhirubhai Ambani Institute of Information 
   and Communication Technology 
    Gandhinagar, Gujarat 
    \\\{201921012, priyanka\_singh\}@daiict.ac.in\\
    \IEEEauthorrefmark{2} Center for Forensic Science,University of Technology Sydney, Australia\\
    \{Manoranjan.Mohanty\}@uts.edu.au}
}

\maketitle

\begin{abstract}
 Photo Response Non-Uniformity (PRNU) noise has proven to be a very effective tool in camera based forensics. It helps to match a photo to the device that clicked it. In today’s scenario, where millions and millions of images are uploaded every hour, it is very easy to compute this unique PRNU pattern from a couple of shared images on social profiles. This endangers the privacy of the camera owner and becomes a cause of major concern for the privacy-aware society. We propose SSS-PRNU scheme that facilitates the forensic investigators to carry out their crime investigation without breaching the privacy of the people. Thus, maintaining a balance between the two. To preserve privacy, extraction of camera fingerprint and PRNU noise for a suspicious image is computed in a trusted execution environment such as ARM TrustZone. After extraction, the sensitive information of camera fingerprint and PRNU noise is distributed into multiple obfuscated shares using Shamir Secret Sharing (SSS) scheme. These shares are information-theoretically secure and leak no information of underlying content. The encrypted information is distributed to multiple third-party servers where correlation is computed on a share basis between the camera fingerprint and the PRNU noise. These partial correlation values are combined together to obtain the final correlation value that becomes the basis for a match decision.Transforming the computation of the correlation value in the encrypted domain and making it well suited for a distributed environment is the main contribution of the paper. Experimental results validate the feasibility of the proposed scheme that provides a secure framework for PRNU based source camera attribution. The security analysis and evaluation of computational and storage overheads are performed to analyze the practical feasibility of the scheme.
\end{abstract}

\section{Introduction}
Photo Response Non-Uniformity (PRNU) is a noise-based source camera attribution technique. The PRNU is based on the noise pattern that is present in a camera sensor due to the manufacturer's flaw. The sensor elements have minute  differences in the area and thus capture different amounts of energy even under a perfectly uniform light condition ~\cite{fernandez2020information}. This PRNU noise pattern is unique to each camera and hence, can be considered as a camera fingerprint. PRNU-based method has potential for a lot of applicaltions in cyber forensics. For instance, it can be helpful to the law enforcement authorities in tracing a cybercriminal based on the camera fingerprint e.g. a imposter sharing child pornography images ~\cite{mohanty2018pandora}. To address this problem, one possible solution could be deducing the PRNU noise from a bunch of shared images on the social media sites and estimating the camera fingerprint. Later, if somebody uploads a child porn image, the PRNU noise can be extracted from it and checked for a possible match with the suspected camera fingerprints ~\cite{mohanty2019prnu}.



The unprecedented flow of millions of images on social media sites has raised alarming privacy concerns as PRNU can be easily computed from a small set of available images, and the identity of a person can be revealed. For example, consider a preliminary investigation of a drug case. A journalist took an image of a drug dealing incident. The identity of the journalist is supposed to be kept confidential to guarantee his life's safety. Now, let us assume that this confidential image is somehow leaked to the public and the drug dealer comes across this image. Based on his past history, he may frame a list of his prime suspects. As the crime's scene image is available, it's PRNU can be easily extracted and based on the suspect's list, he can check for a possible match of the camera. The camera fingerprint can be easily obtained from the
social media sites. At that point, the journalist who has actually clicked that crime scene image is at high risk ~\cite{mohanty2019prnu}. In this whole scenario, PRNU endangered life of a right person. Hence,  preservation of privacy is must for exploiting PRNU to it's advantage. 

To this end, we propose SSS-PRNU that maintains the utility of the PRNU along with preserving the privacy of an individual. 
SSS-PRNU architecture is proposed for a distributed cloud computing environment where the data is outsourced to the third-party cloud servers to avail their services. 
These third party servers are honest but curious so they follow the protocol but still try to obtain information about the underlying content. In SSS-PRNU, we have maintained  confidentiality of the outsourced content along with preserving the utility of PRNU and achieving fault tolerance for the proposed framework. Following are the key contributions:
\begin{itemize}
\item {Confidentiality of outsourced content: The camera fingerprint is computed in a trusted environment and thereafter, it is distributed into multiple random looking shares. These shares are obtained based on Shamir secret sharing (SSS) and hence, are information-theorectically secure. This implies no matter how much computational power an adversary has, the individual shares can never leak any meaningful information.}
\item{SSS-PRNU is a fault-tolerant system: This implies that the utility of the system will be maintained even if some of the servers go down. SSS-PRNU will work unaffected until $2l-1$ cloud servers are up.}
\item{SSS-PRNU is privacy-preserving: In the entire protocol, the privacy of an individual is not compromised. The extraction of camera fingerprint is done in a secure trusted zone and encrypted prior to distribution over third party servers. PRNU-noise of the suspected image is also encrypted using SSS and the possible matching with the camera fingerprint  is done on a share-to-share basis using a correlation metric.}
\item{Computation of correlation in encrypted domain: SSS-PRNU has exploited the homomorphic properties of SSS to compute Pearson correlation coefficient in the encrypted domain. As SSS allows additions and only one multiplication operation in the ciphertext, we have limited to these homomorphic properties for computing coefficient in parts. One multiplication operation is supported in SSS only if we have atleast $2l-1$ shares. So, for SSS-PRNU  to enable computation of correlation in encrypted domain,  we need atleast $2l-1$ shares.}
\end{itemize}



The rest of the paper is organized as follows. Section \ref{preliminaries} presents a brief overview of PRNU-based source camera attribution, SSS scheme, and Trusted Execution Environment (TEE). In section \ref{related}, we review the related work. Section \ref{architecture} describes the architecture of the proposed scheme, the system and the threat model. Section \ref{proposed} details the proposed SSS-PRNU scheme. The solution details and the security analysis are done in section \ref{solutiondetails} and \ref{security} respectively. Section \ref{experiment} describes the experiments conducted to validate the performance of the proposed scheme and section \ref{conclusion} concludes the work along with future scope.

\section{preliminaries}
\label{preliminaries}
In this section we describe the PRNU-Based Source Camera Attribution.
\begin{figure*}[h!]
    \centering
    \includegraphics[scale=0.85]{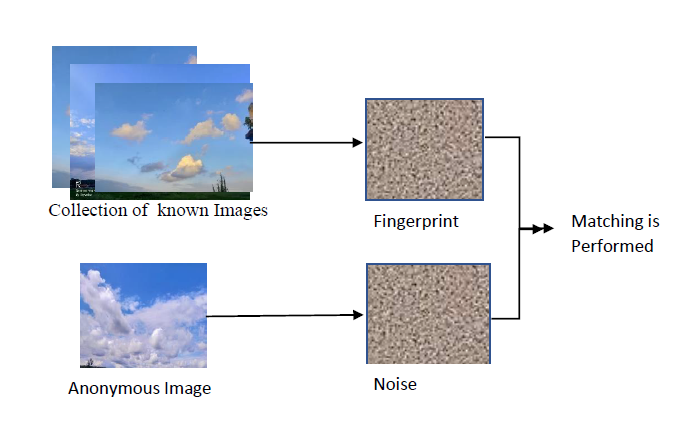}
    \caption{PRNU:A fingerprint is caluculated from a set of known images and noise extracted from the anonymous image. Matching of fingerprint is performed to check whether an anonymous image is taken from the same camera.}
    \label{PRNU}
\end{figure*}

\subsection{PRNU-Based Source camera attribution}

The Photo Response Non-Uniformity (PRNU) is an inherent
properties of each camera, caused by slight variations in the manufacturing process of the sensor. It is based on the sensor output $L$ from a camera which can be modeled as\cite{taspinar2017prnu} :
\begin{equation}
   L = L^{(0)} + L^{(0)}X + \xi,
\end{equation}
Where $L^{(0)}$ is a noise-free result, $X$ is the PRNU noise, and $\xi$ is a combination of additional noise which is considered for all sorts of disturbances. Since $\xi$ is a random noise they cannot be removed by using a denoising filter. Therefore, multiple still images are averaged to minimize and improve the estimation of X, which is PRNU noise 
called the camera fingerprint.
 We take a collection of images from the camera be denoted as 
 $L_{k}$ where, $k = 1,2,\dots,n$. Then take a filtered image $ \vartheta(L_{k})$ using a denoising filter $\vartheta$ (wavelet filter).
 Camera fingerprint is calculated for the $k^{th}$ image:
\begin{equation}
    X_{k} = L_{k} - \vartheta(L_{k})
\end{equation}
Calculate the average by combining the camera fingerprint $X$ of all the images. 
Further, take the query image $L^{'}$, we calculate the PRNU 
noise($X^{'}$) of a query image using denoising filter  
$\vartheta$, and correlate with camera fingerprint $X$. To determine whether the suspected camera has taken the query image $L^{'}$. If it is nearly correlated, we can say the query image is taken from the suspected camera. Pearson correlation coefficient ($r$) that is given as:
\begin{equation}
\label{Pearson}
   r(X,X^{'}) = \frac{\sum_{k=1}^{n}(X_{k}-\bar{X})(X^{'}_{k}-\bar{X^{'}})}{\sqrt{\sum_{k=1}^{n}(X_{k}-\bar{X})^{2}\sum_{k=1}^{n}(X^{'}_{k}-\bar{X^{'}})^{2}}}\end{equation}

Where $\bar{X}$ = $\frac{\sum_{k=1}^{n}X_{k} }{n}$ and $\bar{X^{'}}$ = $\frac{\sum_{k=1}^{n}X^{'}_{k} }{n}$. If the correlation value is equal to or above the threshold, we can conclude that the query image is taken from the suspected camera.

\subsection{Shamir Secret Sharing (SSS)}
Shamir secret sharing (SSS) is also known as $(l, n)$ threshold scheme or Lagrange interpolation scheme.  In this scheme, the secret $b_{0}$ is divided into $n$ random looking shares. The shares individually reveal no information and have information theoretic security. 

\begin{equation}
\label{SSS}
    G(u) = b_{0} + \sum_{i=1}^{l-1} b_{i} u^{i}mod p
\end{equation}

where, $b_{0}$ is the secret, $p$ is a large prime number and  $b_{1},b_{2},\ldots b_{l-1} $ are the coefficients such that $b_{i} < p$.\\

To recover the secret $b_{0}$, it require atleast $l$ shares. The $l$ distinct share numbers $u_{0,}u_{1},u_{2},\ldots u_{l-1}$ and the corresponding $l$ shares are $y_{0}$, $y_{1}$, $y_{2}$, \ldots, $y_{l-1}$ such that $y_{i}= G(u_{i})$, where  $i = 1,2, \ldots,n$ and reconstructed using $(l-1)$ degree of Lagrange interpolation polynomial $F(u)$ defined as:


\begin{equation}
\label{recover}
F(u) = \sum_{i=0}^{l-1}y_{i} \ s_{i}(u) \ mod p
\end{equation}

where, $s_{i}(u)$ is defined as
\begin{equation}
s_{i}(u){=} \prod_{j = 0,j\neq i}^{l-1} \frac{u - u_j}{u_i-u_j} 
\end{equation}
is the Lagrange function. Solving $F(u)$ which is equivalent to the polynomial $G(u)$ by unisolvence theorem.\\

SSS is homomorphic to addition and  
scalar multiplication. This implies that if we do these operations on ciphertexts and decrypt it, it will result into same value as done in the plaintext domain. Mathematically, it can be depicted as follows ~\cite{6466843}:\\

\begin{equation}
  [b_{0} + a\: mod\: p]_{i}^p = [b_{0}]_i^p +a\:mod\:p 
\end{equation}

\begin{equation}
 [ab_{0}\:mod\:p]_{i}^{p} 
 =a[b_{0}]_i^p\:mod\:p
\end{equation}
where, $b_{0}$ is the secret, p is a large prime number and a is the constant.

SSS can support one multiplication in encrypted domain for a specific case when atleast $(2l-1)$  shares are needed to recover the secret ~\cite{gennaro1998simplified}. 

\subsection{TEE (Trusted Execution Enviroment)}

Increasing the use of software tools in recent years has gradually led to deal with more security issues. Therefore, the processing must be done in the trust zone for increasing the alarm for privacy. The trust zone executes a secure environment and proposes a high security level for various applications.

ARM TrustZone is an example of a trust zone. The ARM TrustZone is a trusted environment that provides security to the implementation of various applications. The architecture of our model is executed in ARM TrustZone. The processor is divided into two cores a secure world and a normal world. Both the worlds are separated by hardware and software extensions. The secure world ensures the susceptible application is executed in the trusted operating systems whereas the application which does not require security can be operated in the normal world. A set of applications in a device helps to take an image and extract a camera fingerprint without any fingerprint tampering. Different forensics techniques are there which can detect the tampering of the pictures~\cite{farid2009image}.
Whereas the normal world includes the guest operating system. while taking the images, it can be tampered leading to camera fingerprint extraction tampering. Camera Fingerprint extraction should always be carried in a trusted environment.

\section{Related Work}
 \label{related}

In this section, we survey schemes that have been proposed based on PRNU. With the objective of maintaining privacy of image owners,  misaligning or weakening PRNU noise by applying strong signal processing tools. Gloe et al. proposed two techniques: (i) applying an undetectable resampling activity to the image, and (ii) forging image inception by interchanging PRNU noise of one camera with another ~\cite{gloe2007can}. Karakucuk et al. stated two adaptive PRNU denoising techniques that iteratively extract PRNU noise from an image based on an evaluated gain factor ~\cite{karakuccuk2015adaptive, dirik2014forensic}.

Privacy can also be maintained for PRNU by misalignment. A self-evident process for accomplishing this misalignment is by applying geometric changes like resizing, cropping, etc. ~\cite{taspinar2017prnu}. There is an effective attack design such as forced seam-carving ~\cite{bayram2013seam}, Patch-based desynchronization ~\cite{entrieri2016patch}, and image stitching that are used for irreversible transformation.

However, weakening PRNU noise and misalignment of fingerprint proved inefficient in preserving privacy as it can resist the operations like signal processing operation, compression, denoising, etc.~\cite{lukas2006digital, rosenfeld2009study}. According to Rosenfeld et al. ~\cite{rosenfeld2009study}, the correlation is significant even after eight rounds of denoising between the noise pattern of an image and the fingerprint of the camera. Misalignment of the fingerprint cannot resist the process of irreversible transformation.  Forced seam carving is a process used for irreversibility ~\ cite{taspinar2017prnu}. Due to the presence of uncarved block in the forced seam images, it becomes difficult to determine the owner of the camera. Dirik et al. ~\cite{dirik2014analysis} proved that the method is inefficient if the uncarved blocks are less than $50 \times 50$ in dimension.
Pedrouzo-Ulloa et al.~\cite{pedrouzo2019efficient} proposed the solution for the privacy of PRNU-based camera attribution where the architecture are implemented using the homomorphically RLWE(Ring Learning with Errors)-based cryptosystem combined with a pre-/post-coding which does not require access to a trusted environment.

Mohanty et al.~\cite{mohanty2019prnu} proposed an approach for the privacy concern of PRNU-based camera attribution through encrypting the camera fingerprint of set of images and PRNU noise of anonymous image and performing the matching operations in an encrypted domain. Using the homomorphically BGN cryptosystem the correlation test is evaluated to see whether the anonymous image is taken from suspected camera. The use of the BGN cryptosystem introduces a high overhead. Therefore, to reduce the high overhead, they used a camera fingerprint digest and PRNU noise digest by excluding some value . This improves efficiency but also reduces the performance. In our paper, we are using Shamir's $(l, n)$ secret sharing scheme. SSS is a keyless threshold scheme. The secret is obfuscated into n shares. For the reconstruction of the secret, we require a minimum of $l$ shares. If there are less than $l$ shares then it is not possible to learn about the secret. Using SSS for performing the matching of PRNU- based technique improves in terms of privacy. SSS is fault-tolerant which gives a major advantage. Consider if n is total number of cloud servers, we require a minimum of $2l-1$ servers for system to work, even if $n-(2l-1)$ cloud servers are unable to participate still the system will work. In our paper, we have used the camera fingerprint and PRNU noise instead of camera fingerprint and PRNU noise digest for performing matching in the encrypted domain.

In this paper, we propose SSS-PRNU, a method that supports the privacy of PRNU based camera attribution. The camera fingerprint and the PRNU noise of the suspected image is distributed into random looking shares using SSS. The encoded shares reveal no information about the underlying content to the cloud servers. As the content is encoded, it can be securely disseminated over the  servers. Our method allows unlimited additions and scalar multiplication and it allows one multiplication with a condition of a minimum of $2l-1$ shares to recover the secret. In the cloud servers, the corresponding shares of PRNU noise and camera Fingerprint are multiplied once for calculating the partial Pearson correlation coefficient. Further, reconstructing the partial correlation coefficient values to get a decrypted values. Using the decrypted value we can compute the Pearson correlation coefficient to check whether the suspected camera has taken the query the image.


\section{Architecture of SSS-PRNU}
\label{architecture}
In this section, we give an overview of SSS-PRNU architecture as depicted in Figure\ref{architecture}, the entities involved and the threat model.   
\begin{figure*}[h!]
	\centering
	\includegraphics[scale=0.85]{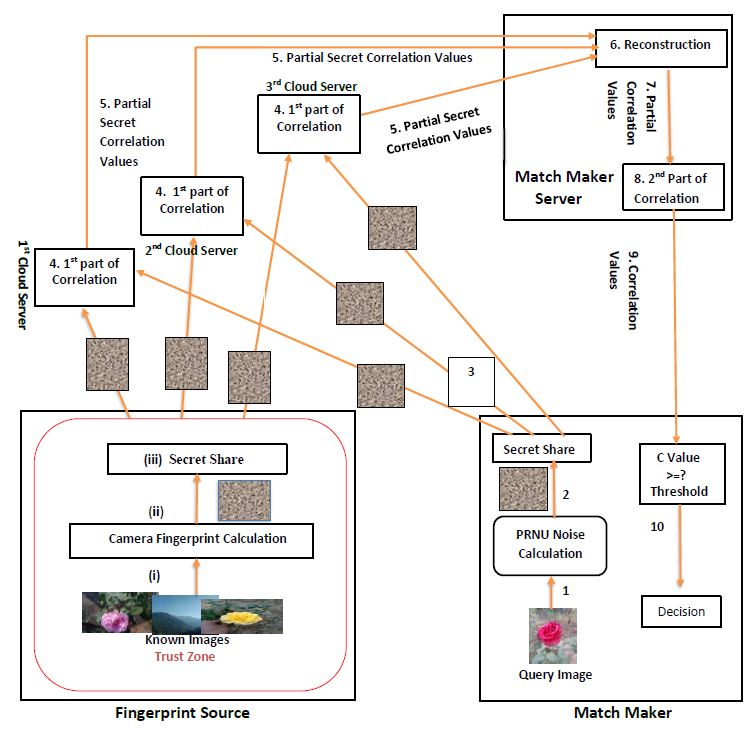}
	\caption{Overview of the Proposed Methodology}
	\label{architecture}
\end{figure*}

\subsection{System Model}
The SSS-PRNU architecture is proposed for a scenario where the data is outsourced in encrypted form to third-party cloud 
servers. It is proposed for a distributed cloud computing 
environment, where instead of outsourcing everything to one 
server, data is distributed over multiple servers. The forensic expert has access to these encrypted files and finds a possible match for an anonymous query image. The matching is done on top of the encrypted data. Thus, preserving the privacy in the entire process. The proposed architecture involves the following entities:\\

\begin{itemize}
\item \textbf{Fingerprint Source}:  A fingerprint source is the entity responsible for computing an authentic camera fingerprint from a set of known images. Authentic camera fingerprint implies that while computing the camera fingerprint, it should be done in a secure environment without any tampers. A tampered camera fingerprint is probable to give high false alarm rates and deteriorate the whole purpose of this methodology. Hence, it is computed in a trusted environment like the ARM trust zone. 
  
To preserve the privacy of the camera owner, the camera fingerprint is encrypted into multiple random-looking shares before it is outsourced to the cloud servers. These cloud servers are third-party and assumed to be honest but curious. They follow the proposed protocol but are also curious to know the underlying content of these shares. But as the shares are generated using SSS, they are information-theoretically secure
and reveal no information. \\

\item \textbf{Cloud Servers as Third-Party Experts}: The cloud servers are the third-party experts where matching of camera fingerprint for a query image is done on a share-basis. The matching is done in the encrypted domain at these cloud servers
and thus, they have no information about the camera or the owner of the camera. They are expected to maintain a database of encrypted shares of the camera fingerprint and have enough computational resources to perform the matching on top of encrypted data. \\

\item \textbf{Match Maker}: This entity is assumed to be a trusted entity e.g., it can be the judge or the law enforcement authority. It has the query image and extracts the PRNU noise from it. After extraction of noise, it obfuscates the information into random shares using SSS scheme prior to outsourcing the content to the third-party cloud servers. 
    The final decision of a suitable match of the extracted PRNU noise with the camera fingerprints is taken by the match maker only.\\

\item \textbf{Match Maker Server}: The Match Maker server receives  a threshold number of shares from the distributed cloud  servers and recovers the partial correlation values  using Lagrange's interpolation. After reconstructing the partial values, it computes the correlation value that is sent to the Match Maker to make the final decision. The Match Maker Server has a smaller infrastructure, storage and computational resources compared to the third party cloud servers and works under the supervision of the Match Maker. 
\end{itemize}

\subsection{Threat model}

 The fingerprint source and the matchmaker server outsource 
 the shares of camera fingerprint and shares of PRNU noise respectively to the cloud server where the processing of operations are implemented. A cloud server is honest but curious. However, the privacy associated with such schemes may get violated since the cloud server might not deviate from the protocol but may try to learn all the possible information which may increase the risk of data privacy. To address this critical problem, it is necessary to enhance the security of data accessing ~\cite{wang2014security}. Shamir secret sharing is an efficient secret sharing algorithm that splits the fingerprint into $n$ pieces at the sender, and the receiver needs to pick up $2l-1$ shares out of $n$ pieces to recover the secret data. The attackers would need to spend effort in getting enough shares to recover the data. An adversary who holds up to $2l-2$ shares in the encrypted domain removes the threat of any data leakage. The security is strengthened by working on shares and processing in the protected domain.
 Hence, the security of data access could be improved by adopting the Shamir secret sharing algorithm. Fingerprint Source, the Match Maker, the Match Maker Server are whole to be trusted zones.

\section{Proposed Methodology}
\label{proposed}
In this section, we describe in detail each of the steps of the
proposed architecture as depicted in Fig. 2. There are three trusted entities: Match Maker, Match Maker Server, Fingerprint Source and the other honest but curious entities are the distributed cloud servers or the third-party experts. The proposed methodology can be adopted to check whether an anonymous query image is taken from a suspected camera or not. The details are as follows:\\

\textbf{Fingerprint Source}\\
Step i: A set of images clicked from a camera device is collected.\\
Step ii: A camera fingerprint is extracted from these images. \\
Step iii: The camera fingerprint in encrypted to obtain multiple random-looking shares using SSS scheme.\\

It is assumed that the camera fingerprints are precomputed and stored in a central database prior to starting the actual protocol.   

\textbf{Match Maker:}\\
Step 1: Extract the PRNU noise from the query image.\\
Step 2: The extracted PRNU noise is encrypted into multiple random-looking shares using SSS. \\
Step 3: The encrypted shares are then distributed to the  third-party experts or the cloud servers.\\

\textbf{Third-Party Experts/Cloud Servers:}\\
Step 4: The cloud servers compute a part of partial correlation coefficient between the share of the encrypted PRNU noise for the query image and the corresponding share of the camera fingerprint received from the Fingerprint Source entity. \\
Step 5: A threshold number $2l-1$ shares of the part of the partial correlation values are collected from the distributed cloud servers and fetched to the Match Maker Server.\\

\textbf{Match Maker Server:}\\
Step 6: The matchmaker server combines the shares of the first part of the correlation values and obtains it's value.\\
Step 7: The reconstruction value is said to be the partial correlation value and it is retrieved by the matchmaker server. \\
Step 8: The second part of correlation is computed using the partial correlation coefficient, and it operates on the remaining operations of the Pearson correlation coefficients.\\
Step 9: The correlation value is computed and sent to the Match Maker.\\

\textbf{Match Maker:}\\
Step 10: The final decision is made by the match maker based on the correlation value it receives from the match maker server.  If the correlation value is equal or above a  specific threshold value, the query image is validated to be taken from the suspected camera else not.

\section{Solution Details}
\label{solutiondetails}
 The camera fingerprint is computed using a  set of images clicked from the same camera. Let us assume the camera fingerprint vector $X$ is represented as $X$ = 
 {$v_1$, $v_2$, $\ldots$, $v_n$}, where $n$ represents the length of the vector. To preserve the privacy of the camera owner, this vector $X$ is encrypted using SSS to distribute the information into multiple obfuscated shares. As the vector contains values that are floating-point, it needs to be scaled and rounded to a nearest integer value as only integers can be encrypted. 
 For scaling, we round off to d decimal places and multiply these floating-point numbers with $10^{d}$ to obtain an integer value. 

 Once we obtain the integer values, it is distributed into multiple shares using SSS scheme using Equation \ref{SSS}. To check for a match of a suspected query image, the PRNU noise$(X^{'})$ is computed for the query image and it is also scaled to obtain an integer value and encrypted using SSS to obtain multiple shares. 
 
Let the encrypted camera fingerprint and the encrypted PRNU noise of the query image be represented as $E(X)$ and  $E(X^{'})$ respectively. To check a match, the Pearson correlation coefficient needs to be computed between these encrypted values on a share basis. This implies at every cloud server,  the partial values for the Pearson correlation coefficient is computed. The Pearson correlation coefficient is computed using Equation \ref{Pearson} and it involves additions, multiplications, a division, and a square root operation. As this correlation is to be computed for encrypted shares of camera fingerprint and PRNU noise, it has be computed in parts as it is limited by the homomorphic properties supported by SSS scheme. In general,  SSS encryption is homomorphic to additions and scalar multiplications. For a special case, when the threshold for the minimum number of shares to recover the secret is set to $2l-1$, it supports  one multiplication too in the encrypted domain. However, it does not support division and square root operation and hence, the need arises to compute it in parts. At each cloud server $i$, the Pearson correlation coefficient is calculated partially using encrypted camera fingerprint share $E(X^{i})$ and the encrypted share of PRNU noise $E(X^{i'})$ to obtain $E(P^{i})$, $E(Q^{i})$, and $E(R^{i})$ for each of these shares in the encrypted domain. The operations involved in  computation of $E(P^{i})$, $E(Q^{i})$, and $E(R^{i})$ are multiplications and additions. The partial components of the Pearson Correlation Coefficient are computed as follows:

\begin{equation}
\label{P_value}
E(P^{i}) = \sum_{k=1}^{n}(E(X_{k}^{i}) - \overline{E(X^{i})})(E(X_{k}^{i'}) - \overline{E(X^{i'})})
\end{equation}

\begin{equation}
\label{Q_value}
    E(Q^{i}) = \sum_{k=1}^{n}(E(X_{k}^{i}) - \overline{E(X^{i})})^{2}
\end{equation}

\begin{equation}
\label{R_value}
    E(R^{i}) = \sum_{k=1}^{n}(E(X_{k}^{i'}) - \overline{E(X^{i'})})^{2}
\end{equation}
   
         where $\overline{E(X^{i})}$ and 
         $\overline{E(X^{i'})}$ represent the mean of $E(X^{i})$ and $E(X^{i'})$
 respectively.\\
 
 To compute the mean values, the operations involved are additions and one scalar multiplication that is supported by SSS scheme. 

 The cloud servers or the third-party experts send the partial encrypted components $E(P^{i})$, $E(Q^{i})$, and $E(R^{i})$ to the 
 matchmaker server. The partial encrypted 
 components are combined together using Lagrange's interpolation to reconstruct the decrypted values $P$, $Q$, and $R$ using Equation \ref{recover}. These values are the scaled ones, so they are descaled by by dividing with $10^d$ and converted back to float values. As we have obtained the plaintext values for $P$, $Q$, and $R$, further computation involving the square root and the division operations can be performed to obtain the final value of Pearson correlation 
 coefficient in the plain text-domain as follows: 
 \begin{equation}
 \label{correlation}
     r(X,X^{'}) = \frac {P}{\sqrt{QR}}
 \end{equation}
 
 The matchmaker takes the decision based on the final correlation value. If the correlation value is greater or equal to the threshold value, the query image is declared as a match and taken from the suspected camera otherwise not.\\
 
  Example $1$: Compute Pearson correlation coefficient between camera fingerprint $X$ and PRNU noise $X^{'}$ in the plaintext domain.
 \begin{figure*}[h!]
    \centering
    \includegraphics[scale=0.32]{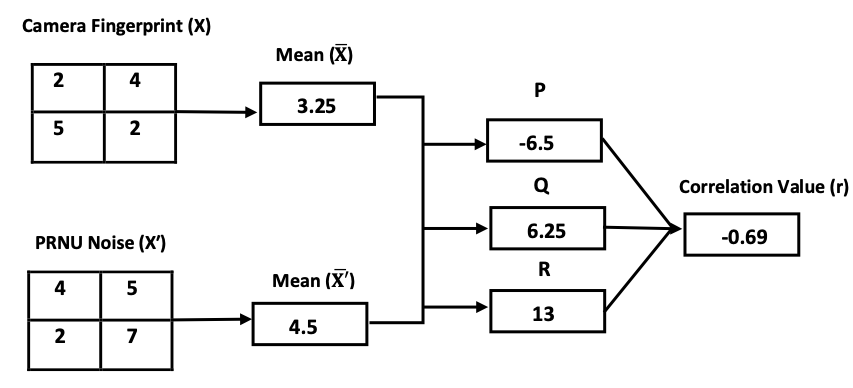}
    \caption{Correlation value computation in plaintext domain}
    \label{plaintext}
\end{figure*}
 
In Figure \ref{plaintext}, the camera fingerprint $X$ and PRNU
noise $X^{'}$ are represented as two $2\times2$ matrices. The 
Pearson correlation coefficient value $r$ is determined using 
equation\ref{Pearson} in the plaintext domain. This correlation value between $X$ and $X^{'}$ is used to verify whether a query image is a match or not to the camera.\\
 
Example $2$: Compute Pearson correlation coefficient between camera fingerprint $X$ and PRNU noise $X^{'}$ in the encrypted domain.\\

 \begin{figure*}[h!]
    \centering
    \includegraphics[scale=0.20]{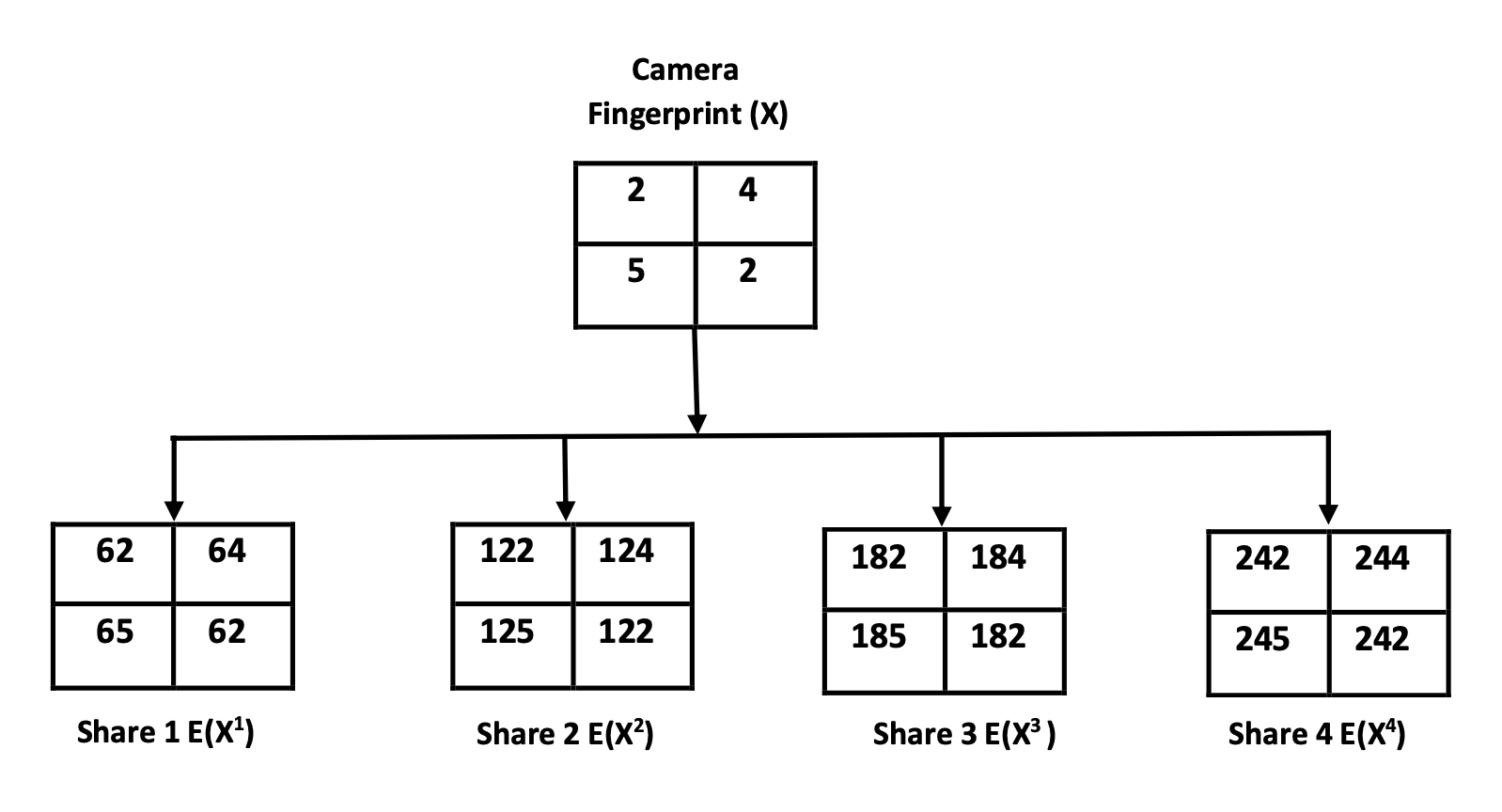}
    \caption{Camera fingerprint obfuscation into multiple shares using SSS scheme }
    \label{matrix_1}
\end{figure*}
 
 \begin{figure*}[h!]
    \centering
    \includegraphics[scale=0.20]{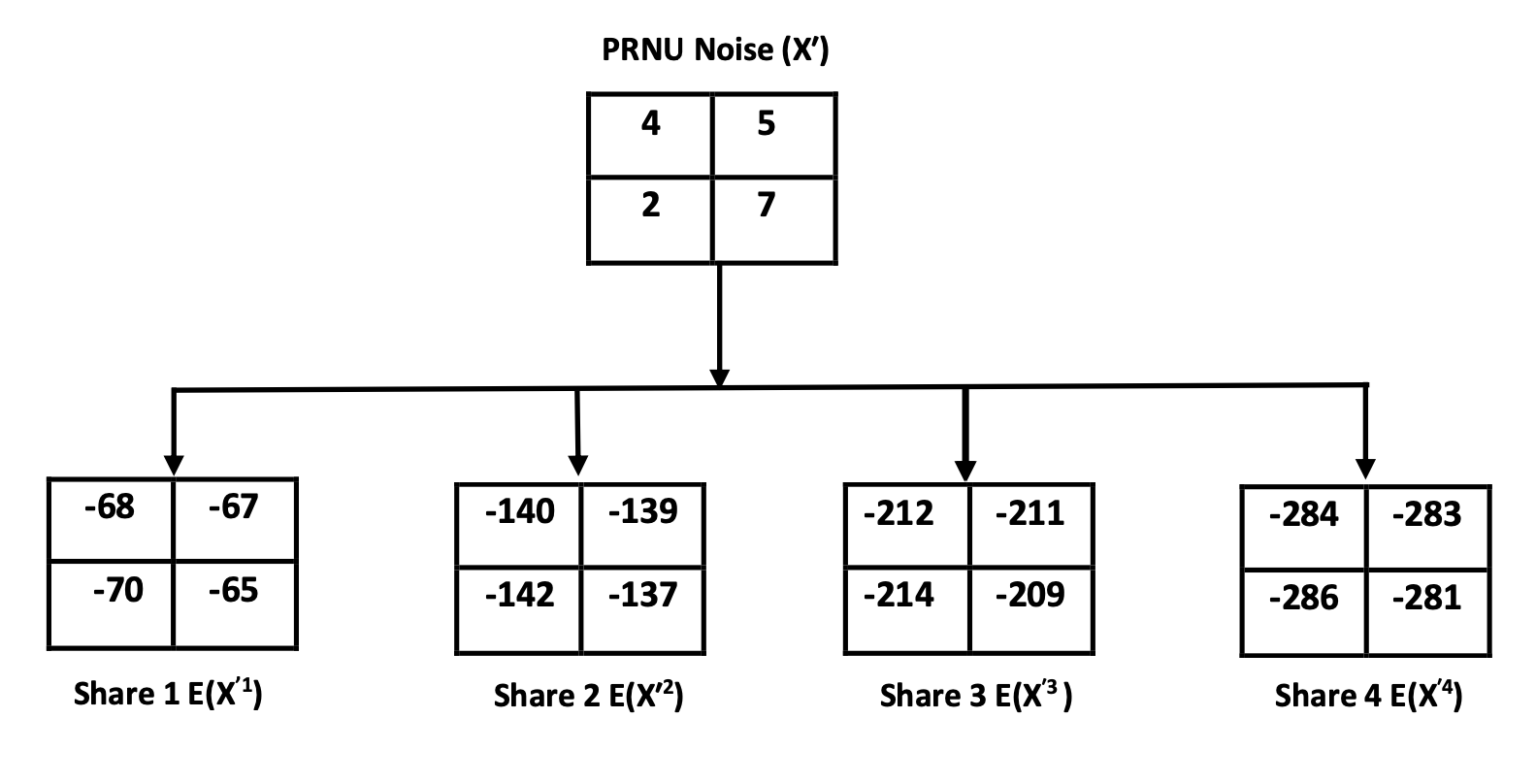}
    \caption{PRNU noise obfuscation into multiple shares using SSS scheme }
    \label{matrix_2}
\end{figure*}

The camera fingerprint $X$ and the PRNU noise $X^{'}$ are obfuscated into random looking shares based on the $(2,4)$ threshold SSS scheme as shown in the Figure \ref{matrix_1} and Figure \ref{matrix_2} respectively.

Further, the correlation value is computed on share basis at the third-party cloud servers. The share of the encrypted camera fingerprint and encrypted PRNU noise present at the $i^{th}$ cloud server are represented as $E(X^{i})$ and $E(X^{'i})$, where $ i= 1,2,\dots n$. The partial correlation coefficient $E(P^{i})$, $E(Q^{i})$, and  $E(R^{i})$ are computed at each $i^{th}$ cloud server using equation \ref{P_value}, \ref{Q_value}, and \ref{R_value}.

The Shares $E(X^{i})$ and $E(X^{'i})$ are multiplied once   and addition operations are involved in computing $E(P^{i})$, $E(Q^{i})$, and  $E(R^{i})$ . We are implementing SSS which is homomorphic to addition, scalar multiplication. It supports one multiplication with a threshold of $2l-1$ to recover the secret. Therefore, it is possible to perform in the encrypted domain.

 \begin{figure*}[h!]
    \centering
    \includegraphics[scale=0.35]{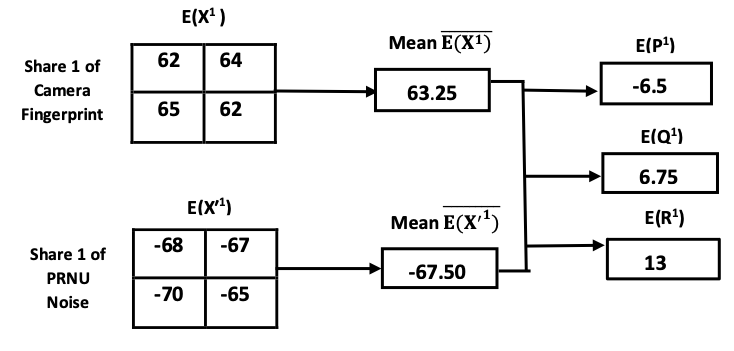}
    \caption{Partial correlation values on share basis}
    \label{cloudserver}
\end{figure*}
 
 \begin{figure*}[h!]
    \centering
    \includegraphics[scale=0.35]{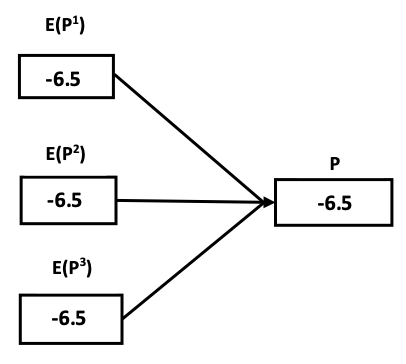}
    \caption{Reconstruction of partial values for correlation using Lagrange's interpolation}
    \label{reconstruction}
\end{figure*}

A minimum $2l-1$ number of shares at the cloud servers send their values of $E(P^{i})$, $E(Q^{i})$, and  $E(R^{i})$ to the match maker server to reconstruct the $P$, $Q$ and $R$  using Lagrange's interpolation as shown in equation \ref{recover}. Once we reconstruct the decrypted values of $P$, $Q$ and $R$, the correlation value $(r)$ is computed using equation \ref{correlation}.

\section{\textbf{Security Analysis}}
\label{security}
The proposed SSS-PRNU scheme provides data confidentiality, data integrity, and data availability.
\subsection{Data Confidentiality}
In SSS-PRNU, fingerprint source, Match Maker and Match Maker Server are assumed to be trusted entities. The Cloud Servers are considered to be honest-but-curious entities. Thus any data leak can happen either from a Cloud Server-end or when data is in transit.\\

\textbf{Lemma 1:} SSS-PRNU is a perfectly secure scheme.\\

$\textbf{Proof(Sketch).}$ SSS-PRNU is based on $(l,n)$ Shamir's secret sharing. 

Shamir's secret sharing is a perfectly secure scheme ~\cite{shingu2016secrecy}. In the case of Shamir's secret sharing, polynomial values (i.e., shares) from $l$ cloud servers are required to reconstruct the $(l-1)$-degree original polynomial and get the secret. 

In the case of SSS-PRNU, polynomial values (i.e., shares) from $(2l-1)$ cloud servers, however, are required to get the secret. This is because SSS-PRNU supports one multiplication. When two values of two $(l-1)$-degree polynomials are multiplied, a value of a $(2l-1)$-degree polynomial is formed. To reconstruct the $(2l-1)$-degree polynomial and get the secret (which is the multiplication of two secrets), shares from $(2l-1)$ cloud servers are required. Thus, SSS-PRNU can be considered as $(k, n)$ Shamir's secret sharing, where $k = 2l-1$. Hence, SSS-PRNU is also perfectly secure.

Since SSS-PRNU is a perfectly secure scheme. The secret cannot be leaked from less than $(2l-1)$ cloud servers. With the same argument, an adversary listening (hacking) the communication channels of up to $(2l-1)$ cloud servers cannot also be able to get the secret. Thus, SSS-PRNU supports data confidentiality.

\subsection{Data Integrity}
SSS-PRNU provides data integrity when the total number of cloud servers (i.e., $n$) is more than the minimum number of required cloud servers (i.e., $(2l -1)$).

There are $n \choose (n- (2l-1)) $ different ways of reconstructing a secret from $(2l -1)$ share values obtained from $(2l-1)$ cloud servers (each time different cloud servers will be used). In the ideal case, when none of the share values is tampered with, all the $n \choose (n- (2l-1))$ reconstructed secrets must be the same. However, when at most $n-1$ share values at $n-1$ cloud servers are tampered, all the $(n- (2l-1))$ secrets will not be the same. The secret reconstructed using the non-tampered share will be different than other reconstructed secrets. Thus by comparing the reconstructed secrets, it can be concluded that tampering has happened. However, if all the shares are tampered by obeying the homomorphic property of Shamir's secret sharing, all the reconstructed secrets will be same, and tampering cannot be detected. 

When $n = (n- (2l-1))$, there is only one way of reconstructing the secret. In this case, tampering also cannot be detected. 

\subsection{Data Availability}
SSS-PRNU provides data availability when the total number of cloud servers (i.e., $n$) is more than the minimum number of required cloud servers (i.e., $(2l -1)$). In such a case, the proposed scheme will work even if at most $(n- (2l-1))$ cloud servers are out of service as there is at least one way of getting the secret.

\section{Experimental Analysis and Results}
\label{experiment}
To validate the proposed SSS-PRNU, we conducted experiments on a PC powered by the Intel (R) Core(TM) i3 - 1005G1 CPU @ 1.20 GHz and RAM 8GB. The fingerprint source, Third-Party Expert, matchmaker, and matchmaker server were simulated on different PC's in our lab. The extraction of camera fingerprint and distribution of it's information into obfuscated shares using SSS was done in ARM TrustZone. It is a TEE and has a processor that is divided into two cores, a normal world and a secure world. The fingerprint extraction and encryption was done in the secure world of this TrustZone, installed on a Intel (R) Core(TM) i3 - 1005G1 CPU @ 1.20 GHz and RAM 8GB machine with Ubuntu as the operating system. 

We used MATLAB R2019b to compute the 
camera fingerprint from a set of images, PRNU noise of a query image, and obtain obfuscated shares of both using SSS. In Section 2, we have explained the generation of 
fingerprint and PRNU noise in the plaintext domain. The fingerprint and PRNU noise is a floating-point number, we  first round off the floating point number to d decimal places then multiply $10^{d}$  to round off number to obtain an integer value so that we can encrypt them and transform to the encrypted domain. The SSS scheme is applied to encrypt the fingerprint and the PRNU noise into multiple random looking shares. The minimum number of shares required to recover the encrypted information is set to $2l-1$.  
The threshold is set to $2l-1$ so that the SSS scheme could support one multiplication operation in the encrypted domain. 

For the experiments, we implemented SSS as $(2,4)$ scheme that required atleast $3$ shares to retrieve the secret. 
The fingerprint source encrypts the fingerprint and the matchmaker server does the encryption of the PRNU noise of the query image. The third-party cloud servers compute the correlation between the corresponding shares of the camera fingerprint and the PRNU noise using Pearson correlation coefficient. The encoded shares are information-theoretic secure and reveal no information about the fingerprint and PRNU noise to the cloud servers. As the correlation is computed in the encrypted domain, it is possible to compute only the partial correlation values, supported by the homomorphic properties of the SSS scheme as detailed in section \label{solution}.  The matchmaker reconstructs the  encrypted partial correlation values, computes the final correlation value, and matches the threshold. SSS-PRNU provides a secure framework for extraction of fingerprints and PRNU noise, distribution of extracted information into obfuscated shares using SSS scheme, processing these shares in encrypted domain to obtain the correlation coefficient value, and perform the matching to arrive at a decision.

 We took 12 cameras of 12 different brands for our experimental study, details enlisted in Table~\ref{camera_list}, along with the Pearson correlation coefficient value computed in the plain text domain and the encrypted domain. We can observe from the Table~\ref{camera_list} that the correlation values come out to be same in plaintext as well as in the encrypted domain. Thus, we conclude that the PRNU computation can be done in a privacy preserving manner in the encrypted domain based on SSS scheme. 
\begin{table*}[]
\caption{Values of correlation coefficient for various cameras} 
    \centering
    \begin{tabular}{| m{3 cm} | m{2cm}| m{2cm}| } 
    \hline
     \textbf{Mobile Phone} & \textbf{Plaintext} & 
     \textbf{Encrypted} \\
      & \textbf{Domain} & 
     \textbf{Domain} \\
     \hline
     J7 Galaxy   & 0.5333 & 0.5333\\  
     \hline
     Redmi Note 4   & 0.4493 & 0.4493\\
     \hline
     IPhone SE  & 0.3308 & 0.3308 \\
     \hline
     IPhone 6   & 0.447 & 0.447 \\
     \hline
     One Plus 3T   & 0.3047 & 0.3047 \\
     \hline
     Samsung M31   & 0.0205 & 0.0205 \\
     \hline
     Canon 200   & 0.0156 & 0.0156 \\
     \hline
     A15 Xiaomi 3   & 0.3346 & 0.3346 \\
     \hline
     Gionee   & 0.4121 & 0.4121 \\
     \hline
     Samsung Tab S6 Lite   & 0.0141 & 0.0141 \\
     \hline
     Samsung A50S   & 0.0019 & 0.0019 \\
     \hline
     Samsung Galaxy J3   & 0.0087 & 0.0087 \\
     \hline
    \end{tabular}
    \label{camera_list}
\end{table*}

\begin{table}[]

\caption{Computation time(in seconds) for camera fingerprint extraction and obfuscation into shares }
    \centering
    \begin{tabular}{ | m{3cm} | m{2cm}| m{2cm}| } 
    \hline
     \textbf{Phone} & 
     \textbf{Plaintext} & \textbf{Encrypted} \\
      & \textbf{Domain} & 
     \textbf{Domain} \\
     \hline
     J7 Galaxy  & 0.7033  & 41.6251 \\
     \hline
     Redmi Note 4  & 0.7155 & 52.4346\\
     \hline
     IPhone SE  & 0.9815  & 45.9742\\
     \hline
     IPhone 6  & 1.0882 & 44.1546\\
     \hline
     One Plus 3T & 1.4533  & 42.8338\\
     \hline
     Samsung M31  & 1.4087 & 39.3485 \\
     \hline
     Canon 200 & 1.933  & 42.5231 \\
     \hline
     A14 Xiaomi 3 & 1.3385  & 38.4134 \\
     \hline
     Gionee  & 1.3151 & 32.255\\
     \hline
     Samsung Tab S6 Lite  & 0.99 & 32.7041 \\
     \hline
     Samsung A50S & 1.2241   & 34.4113 \\
     \hline
     Samsung Galaxy J3 & 1.414 & 34.7841 \\
     \hline
    \end{tabular}
    \label{computation_cost_fingerprint}
\end{table}

\begin{table}[]
\caption{Computation time(in seconds) for PRNU noise extraction from the query image and computing correlation coefficient }
    \centering
    \begin{tabular}{ | m{3cm} | m{2cm}| m{2cm}| } 
    \hline
      \textbf{Phone} & 
     \textbf{Plaintext} & \textbf{Encrypted}  \\
      & \textbf{Domain} & 
     \textbf{Domain} \\
     \hline
     J7 Galaxy & 5.5548  & 32.8847 \\
     \hline
     Redmi Note 4 & 3.6077  & 31.9287 \\
     \hline
     IPhone SE & 5.5922   & 30.9356 \\
     \hline
     IPhone 6 & 4.98  & 34.0379 \\
     \hline
     One Plus 3T & 8.55  & 34.402 \\
     \hline
     Samsung M31  & 8.2966 & 35.7813 \\
     \hline
     Canon 200  & 10.8511 & 34.2358 \\
     \hline
     A15 Xiaomi 3 & 7.6443  & 39.1875 \\
     \hline
     Gionee  & 4.1632 & 35.9472\\
     \hline
     Samsung Tab S6 Lite & 6.6552  & 38.1039 \\
     \hline
     Samsung A50S & 6.5872  & 34.7052 \\
     \hline
     Samsung Galaxy J3 & 6.4164   & 34.8359 \\
     \hline
    \end{tabular}
    \label{computation_cost}
\end{table}

\subsection{Performance Analysis}
In this section, we analyze the performance of SSS-PRNU in terms of computational cost and data overhead. The proposed method performs encryption that incurs additional computation cost compared to it's plaintext version. Our aim here is to examine the computation cost and storage overhead.

In SSS-PRNU, the camera fingerprint and the PRNU noise of the query image is encrypted using SSS scheme to obtain multiple obfuscated shares. As camera fingerprint and PRNU noise are floating-point values, they need to be scaled by doing one round-off operation and then multiplying by $10^{d}$ to obtain an integer value. These integer values are then distributed into multiple shares. For camera fingerprint, this is a one-time operation and can be stored in the database. But for encrypting PRNU noise, this has to be done at run time by the match maker. The computational cost depends on the dimensions of the image. 

SSS supports addition, scalar multiplication and one multiplication operation to be done on top of the encrypted content. There operations are equivalent to performing multiplication and exponentiation in the encrypted domain. 
Hence, the Pearson correlation coefficient value is computed in three parts $P$, $Q$, and $R$. At each cloud server, these three values are computed for the corresponding shares of camera fingerprint and PRNU noise. Thus, in comparison to the plaintext domain, the third party cloud servers require more computation resources for computing these values. The match maker server decrypts the values received from the cloud servers by recovering the partial values of the Pearson correlation coefficient by using three Lagrange's interpolation. The minimum number of cloud servers required for this reconstruction is $2l-1$. The decrypted values of the partial values of the Pearson correlation coefficient are combined to obtain the final correlation value that is sent to the match maker for the decision. The computations at the third party, matchmaker, and matchmaker server are performed at run time.

The storage requirement for the SSS-PRNU depends on the $(l,
n)$ threshold scheme. The camera fingerprint say of size $|F|$ bytes is distributed into $n$ shares. The storage requirement for each share is $|F|$ bytes and the total requirement is $n \times |F|$ bytes. These shares are sent to the cloud server. Similarly, PRNU noise is encrypted to multiple shares and sent to the cloud server. 

Camera fingerprint and PRNU noise contain values that are in floating point number. It needs to be scaled and rounded to a nearest integer value as only integers can be encrypted. Each floating-point number is represented as a 32-bit. For scaling, we round off to d decimal places and multiply these floating-point numbers with $10^d$ to obtain an integer value.We create $n$ shares and each share contains integer of size $64$ bits. Hence, the encrypted domain requires $64 \times n$ bits of storage overhead, each for camera fingerprint and PRNU noise. In our implementation, n = 4, and as a result,the data overhead in storing camera fingerprint and sending the encrypted shares of camera fingerprint to the cloud server is increased by 128 times as compared to plaintext domain. Similarly, for the encrypted shares of PRNU noise that are sent to the cloud servers also increase by 128 times.  

For the experiments conducted, we have done the time analysis for various steps in the plaintext domain and in the encrypted domain. Let the time taken for extracting the camera fingerprint from a set of images be denoted as $T_{1}$ and for  obfuscating it into multiple share be $T_{1}^{'}$.  In Table~\ref{computation_cost_fingerprint}, the plaintext domain and encrypted domain columns enlists these $T_{1}$ and $T_{1}^{'}$ values. 

In plaintext domain, let the sum total of the time taken for extraction of PRNU noise from the query image  and for computing the pearson correlation coefficient be denoted as $T_{2}$. To do the same in the encrypted domain, first we will extract the PRNU noise, then obfuscate it into shares and then compute the correlation. Let the sum total of the time taken for obfuscating the PRNU noise into shares and computing the correlation be denoted as $T_{2}^{'}$.
 In table ~\ref{computation_cost}, the values for $T_{2}$ and $T_{2}^{'}$ are enumerated for the plaintext domain and encrypted domain. 
 Here, we are assuming that the camera fingerprints and PRNU noise of the query image are precomputed in a trusted environment and available to the respective entities.

\section{Conclusion and Future Work}
\label{conclusion}
PRNU based camera attribution is a technique widely used for identifying the anonymous query image. In this method, the camera fingerprint is calculated from a set of known images, and PRNU noise is extracted from the query image. Matching of the camera fingerprint and PRNU noise is performed using the Pearson correlation coefficient to check whether an anonymous image is taken from the suspected camera. PRNU- based technique is used broadly in the area of digital forensics, whereas preserving privacy for the user is a major concern. The PRNU-based technique should be only used by the law enforcement authorities in tracing a cybercriminal. Our paper proposed a method to show PRNU based method in the plaintext domain and in the encrypted domain. In our solution the cloud servers do not interact with each other during the computation. This paper encrypts the camera fingerprint and PRNU noise to perform the Pearson correlation in the encrypted domain to check whether the query image is taken from the suspected camera. 

In the future, we can extend this paper by using the Peak-to-Correlation Energy (PCE) instead of Pearson correlation coefficient. In the Pearson correlation co-efficient the threshold value for all the cameras is different whereas PCE is an approach that works on a general threshold for all the cameras, using PCE can lead to reducing the computation cost.

\bibliographystyle{plain}
\bibliography{refn.bib}
\end{document}